\documentclass[twocolumn,notitlepage,tightenlines,preprintnumbers,amsmath,amssymb]{revtex4}

\usepackage{hyperref}
\usepackage{graphicx,amsmath}
\usepackage{epsf}

\begin{document}
\title{Phonon-Polaritons in Nanoscale Waveguides}
\author{Hashem~Zoubi}
\email{hashemz@hit.ac.il}
\affiliation{Department of Physics, Holon Institute of Technology, Holon 5810201, Israel}
\date{09 April, 2018}

\begin{abstract}
We investigate the phenomena of Brillouin induced opacity in nanoscale linear waveguides and Brillouin induced transparency in nanoscale ring waveguides. The concept of phonon-polariton is required in order to get a deep understanding of such phenomena. Phonon-polaritons are coherent superposition of photons and phonons obtained in the strong coupling regime induced through stimulated Brillouin scattering in the presence of a pump field. The formation of phonon-polaritons can be observed via linear optical spectra that are calculated here using the input-output formalism. For a linear waveguide the signature of phonon-polaritons appears as two peaks in the transmission spectrum with an opacity gap, and for a ring waveguide as two dips in the transmission spectrum with a transparency gap.
\end{abstract}

%Keywords: Nanoscale Waveguides, Phonon-Polaritons, Stimulated Brillouin Scattering, Input-Output Formalism, Linear Optical Spectra

\maketitle

\section{Introduction}

Stimulated Brillouin Scattering (SBS), or the scattering of light from mechanical excitations of a medium \cite{Boyd2008,Agrawal2013}, was reported by Brillouin in (1922) but observed till after the invention of the laser by Chaio et al. in (1964) \cite{Chiao1964}. In the recent years big progress took place in fabricating waveguides with nanoscale cross section that opens new horizons for
SBS. A breakthrough has been appeared for SBS in which radiation pressure dominates over the conventional electrostriction mechanism, as predicted in \cite{Rakish2012} theoretically and realized later
experimentally \cite{VanLaer2015a}. Radiation pressure can provide strong photon-phonon interactions that
lead to a significant enhancement of SBS in nanoscale waveguides at relatively
low
power \cite{Zoubi2016}. Such a progress introduces SBS as a promising candidate for quantum information processing involving photons and phonons \cite{Thevenaz2008,Bahl2012,Agarwal2013a,Beugnot2014,Merklein2016,Zoubi2017}.

Several proposals appear in the literature for the realization of nanoscale
waveguides \cite{Eggleton2013}. Among the main
factors that strongly limit the efficiency of each device is the waveguide
mechanical quality factor. For example, in on-chip waveguides the quality factor is
low, due to a direct contact between the waveguide and the substrate
material, and that lead to a relatively fast leak of the phonons into the substrate \cite{Pant2011}. On the
other hand, suspended nanoscale
waveguides, e.g. of silicon nanowires, have higher quality factor with longer phonon
lifetime, but they are limited to a very short waveguide length that yields weak SBS \cite{Shin2013}. A compromise has been suggested by using nanoscale
silicon photonic wires that are supported with a tiny pillar, the fact that keeps a reasonable
quality factor and allows the achievement of relatively long waveguides with strong SBS \cite{VanLaer2015a,VanLaer2015b,Kittlaus2015}.

Controlling the propagation of light in an optical medium can be achieved using
Electromagnetic Induced Transparency (EIT), which can produce fast and slow
light where coherent destructive interference prevents electronic excitations within an optical medium \cite{Fleischhauer2005}. In EIT the phase-matching requirement is automatically
satisfied by the existence of the atomic component. On the other hand EIT has been demonstrated
in cavity optomechanics through the coupling among vibrational modes and photon modes by means of radiation pressure \cite{Safavi-Naeini2011}. Here
the photons and phonons are in-resonator localized and the phase-matching is
also automatically satisfied \cite{Weis2010}. SBS has been suggested for the
generation of fast and slow light, while the short
lifetime of acoustic phonons considered as an obstacle in front of EIT based on
SBS. However, in \cite{Kim2015}
they demonstrate Brillouin scattering induced transparency in exploiting
long-lived propagating light and phonons in a silica resonator under the
required phase-matching.

In the present paper we investigate Brillouin Induced Opacity (BIO) in linear nanoscale
waveguides and Brillouin Induced Transparency (BIT) in ring nanoscale waveguides. The
work is motivated
by the recently observed significant photon-phonon coupling in silicon nanowires \cite{VanLaer2015a}. Phonon coherence lifetimes in
silicon nanowires can easily meet the long-lived coherence that is required for BIT and BIO, where photons and phonons obey
the phase-matching requirement for both forward and backward SBS. In the strong photon-phonon coupling regime, when the coupling parameter is larger than the photon and phonon damping rates, the picture of independent photons and phonons breaks down and the physical excitations are new quasi-particles that termed phonon-polaritons, which are a coherent superposition of photons and phonons.

We consider two types of dielectric waveguides: (i) A linear waveguide that is
supported with two effective mirrors at the two edges. (ii) A ring waveguide that is coupled to an external nanofiber at a fixed point. Photons and phonons can propagate with fixed wavenumbers to the left and the right along the linear waveguide, or clockwise and counter-clockwise in a ring waveguide, hence the momentum-space representation is appropriate in the present context. In applying the input-output formalism, the external-internal field coupling and the leak of photons through the
mirrors in linear waveguides, and through the nanofiber in ring waveguides, are included. The method provides optical spectra, e.g. reflection, transmission and absorption parameters, besides the phase
shifts of the transmitted and reflected fields relative to the incident
ones. The propagating nature of the photons and phonons in waveguides makes
the present discussion significantly different from localized interacting ones inside
a resonator, as for cavity quantum optomechanics \cite{Aspelmeyer2014}. We
investigate the possibility of BIT and BIO in such systems. In the presence of a pump field, we show that an effective coupling
between photons and phonons due to SBS inside the waveguide can result in a complete reflection for an
external signal field in linear waveguides, and a complete transmission in ring waveguides. Namely, at a frequency where a resonance appears between
an external signal field and a cavity photon, a gab opens and the external signal field
is
reflected or transmitted. The phenomena is deeply explained in terms of
phonon-polaritons.

\section{Interacting Photons and Phonons in Nanoscale Waveguides}

\begin{figure}
\includegraphics[width=0.8\linewidth]{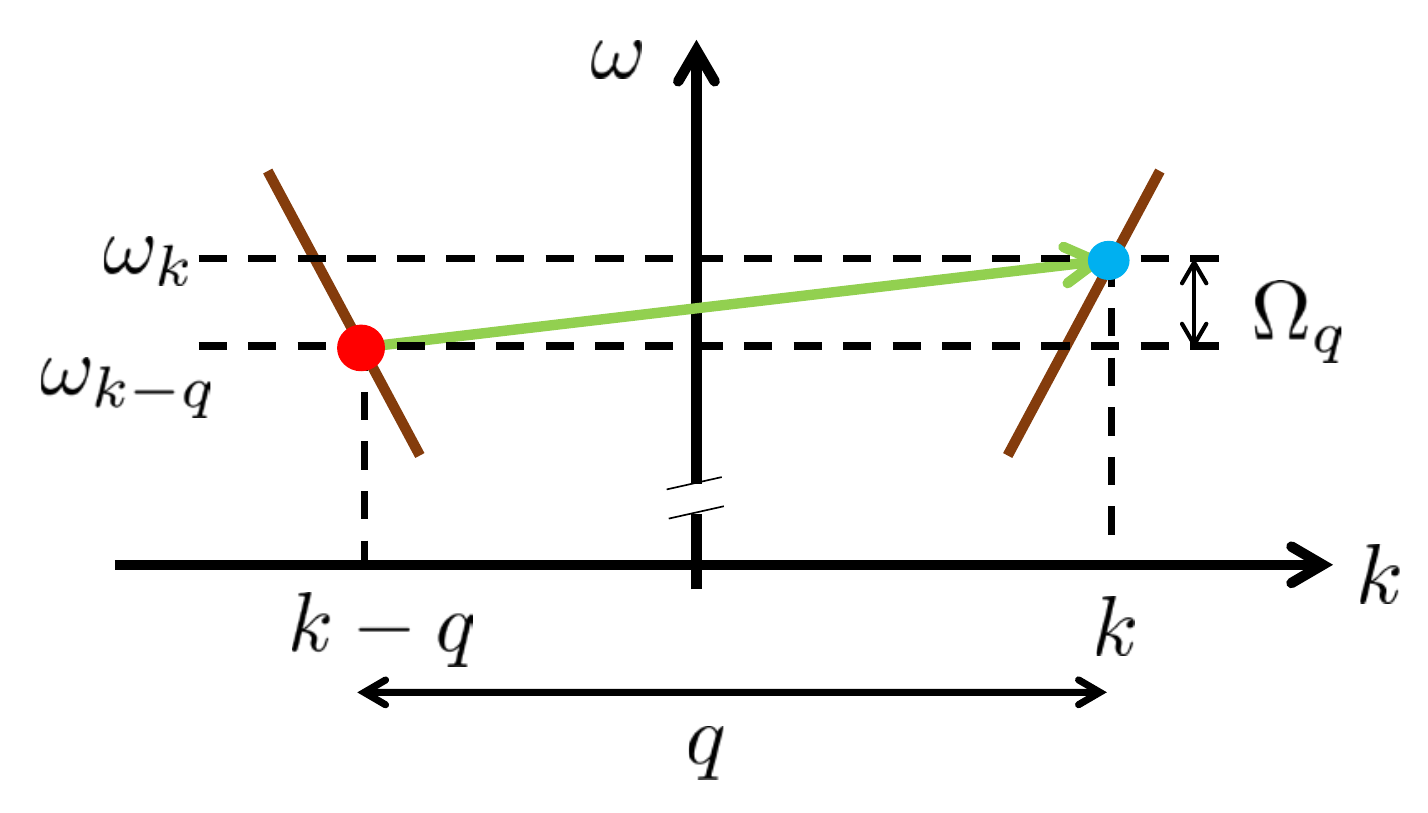}
\caption{Schematic plot of the photon dispersion, $\omega$ vs. $k$, for positive and negative wavenumbers. The phonon dispersion, $\Omega$ vs. $q$, is presented. The process of the scattering of a signal photon, $\omega_k$, into a pump photon, $\omega_{k-q}$, with the emission of a phonon, $\Omega_q$, is plotted.}
\label{PhotPhonDis}
\end{figure}

We present first photons and phonons in nanoscale waveguides where the results hold for both linear and ring waveguides. We consider a nanoscale waveguide made of high contrast dielectric material of
refractive index $n$. The electromagnetic field can freely propagate along the waveguide
axis with almost continuous wavenumbers, for enough long waveguide, and is strongly confined in the transverse direction with discrete modes. Here we
assume a single active mode in the transverse direction. Exploiting translational symmetry, the wavenumber takes the values
$k=\frac{2\pi}{W}p$, where $W$ is the waveguide length with
$(p=0,\pm1,\pm2,\cdots,\pm\infty)$. Moreover, we concentrate in a region where light can propagate with almost
linear dispersion of an effective group velocity of $v_g$. Hence, the photon
dispersion is given by $\omega_k=\omega_0+v_gk$, around a given frequency  $\omega_0$ in the appropriate region. The photon dispersion
is schematically plotted in figure (\ref{PhotPhonDis}). The photon
Hamiltonian in momentum-space representation reads
\begin{equation}
H_{phot}=\sum_k\hbar\omega_k\ a_k^{\dagger}a_k,
\end{equation}
where $a_k$ and $a_k^{\dagger}$ are the creation and annihilation operators of
a photon with wavenumber $k$ for a given transverse mode and a fixed
polarization, with frequency $\omega_k$.

Next we consider the mechanical vibrations in the waveguide. In nanoscale
waveguides different types of vibrational modes can be excited, but here we consider only the propagating
acoustic phonons. The acoustic phonons have a linear
dispersion that is given by $\Omega_q=v_aq$, where $q$ is the acoustic phonon
wavenumber and $v_a$ is the sound velocity. The phonon dispersion is illustrated in
figure (\ref{PhotPhonDis}) for the acoustic mode. The phonon Hamiltonian reads
\begin{equation}
H_{phon}=\sum_{q}\hbar\Omega_{q}\ b_q^{\dagger}b_q,
\end{equation}
where $b_q$ and $b_q^{\dagger}$ are the creation and annihilation operators of
a phonon with wavenumber $q$ for an acoustic mode, of frequencies $\Omega_q$.

The SBS among photons and phonons is subjected to conservation of energy and
momentum. The SBS Hamiltonian is given by \cite{Zoubi2016}
\begin{equation}
H_{SBS}=\hbar\sum_{kq}\left(g_{kq}^{\ast}\ b_q^{\dagger}a_{k-q}^{\dagger}a_k+g_{kq}\ b_qa_{k}^{\dagger}a_{k-q}\right),
\end{equation}
where $g_{kq}$ is the SBS coupling parameter among the two photons of
wavenumbers $k$ and $k-q$ and a phonon of wavenumber $q$. The coupling parameter can be in general $k$ and $q$ dependent, but we assume here the local
field approximation, that is $g_{kq}=g$. The first term represents a process in which a photon, $(k,\omega_k)$, scatters into another photon, $(k-q,\omega_{k-q})$, with the emission of a
phonon, $(q,\Omega_q)$, under the conservation of energy,
$\omega_k-\omega_{k-q}=\Omega_q$. The second term represent a photon, $(k-q,\omega_{k-q})$, that absorbs a phonon, $(q,\Omega_q)$, and scatters into another photon, $(k,\omega_k)$, with the conservation of energy,
$\omega_k-\omega_{k-q}=\Omega_q$. The process is represented schematically in figure (\ref{PhotPhonDis}).

\begin{figure}
\includegraphics[width=0.8\linewidth]{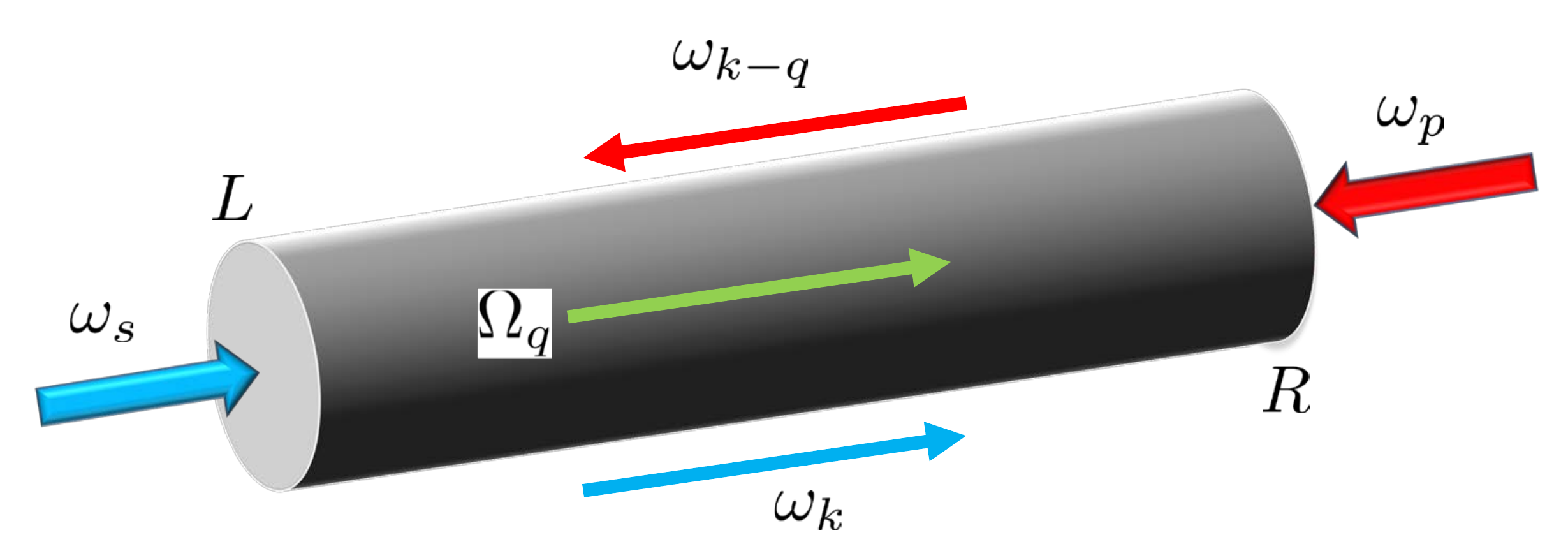}
\caption{The linear nanoscale waveguide in which the two sides are considered as two effective mirrors $L$ and
  $R$, and which allow coupling among external radiation field and internal waveguide field. Inside the waveguide three fields are represented, the photons $\omega_k$, $\omega_{k-q}$ and the phonon $\Omega_q$. Two external fields also appear, the pump field, $\omega_p$, on the right, and the signal field, $\omega_s$, on the left.}
\label{Linear}
\end{figure}

\begin{figure}
\includegraphics[width=0.8\linewidth]{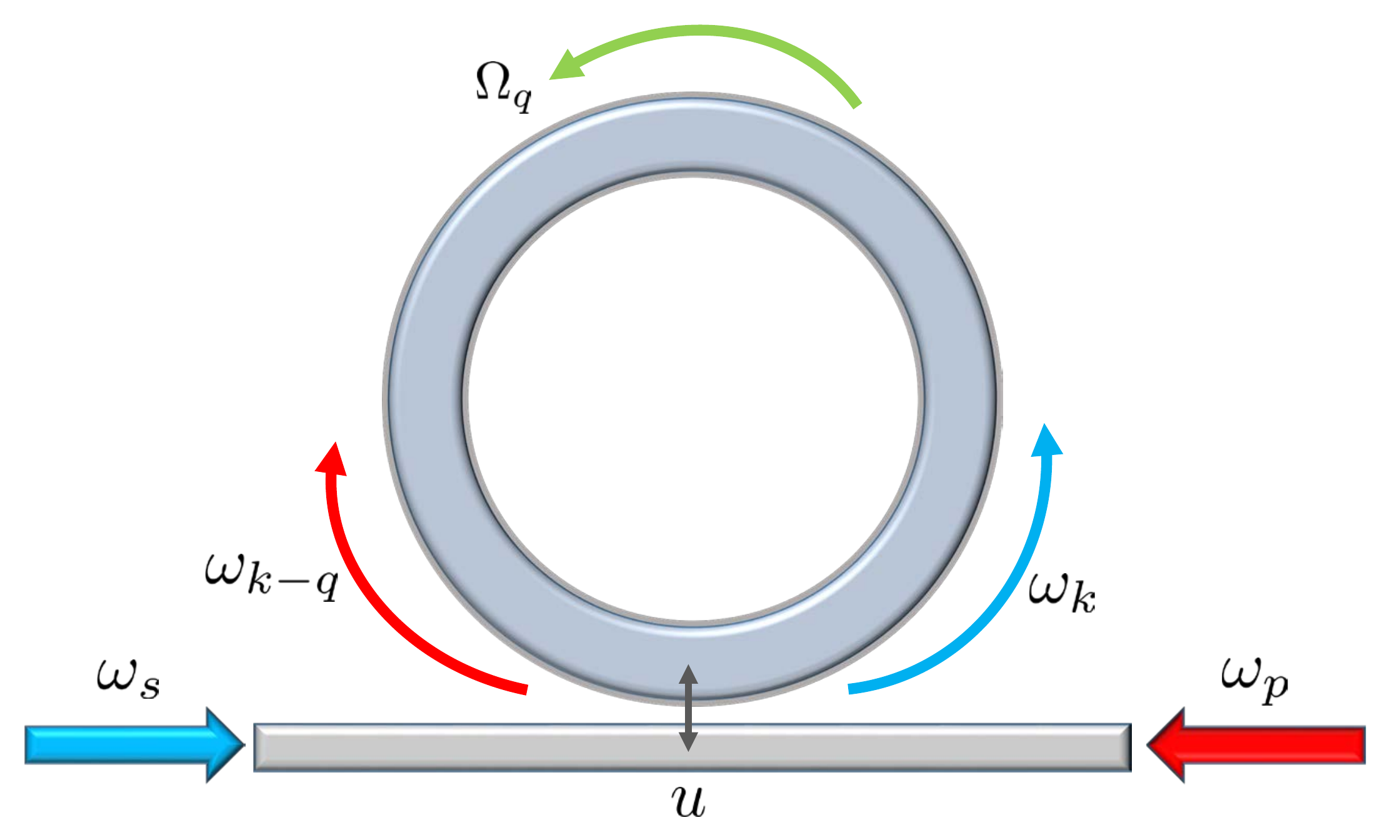}
\caption{The ring nanoscale waveguide that is coupled to an external nanofiber with coupling parameter $u$ that allows tunneling of photons in and out of the waveguide. Inside the waveguide three fields are represented, the photons $\omega_k$, $\omega_{k-q}$ and the phonon $\Omega_q$. Two external fields also appear on the two sides of the external nanofiber, the pump field, $\omega_p$, and the signal field, $\omega_s$.}
\label{Ring}
\end{figure}

In order to excite the waveguide photons one need to couple the internal field to the external radiation field in using the input-output formalism \cite{Walls2008,Gardiner2010}. We consider first the linear waveguide of a finite length and we assume two effective mirrors at the two edges that are denoted by $L$ and $R$, as seen in figure (\ref{Linear}). The mirrors serve us to couple
the external field with the internal waveguide field, and also photons can leak
outside the waveguide through the mirrors. The input and output external fields
of wavenumber $k$, from mirror $L$ are given by $c_{kL}^{in}$ and
$c_{kL}^{out}$, and from mirror $R$ are given by $c_{kR}^{in}$ and
$c_{kR}^{out}$, and which are related to the waveguide mode $k$ by the boundary
conditions \cite{Walls2008,Gardiner2010}
\begin{equation}
c_{kL}^{in}+c_{kL}^{out}=\sqrt{u_{kL}}a_k,\ c_{kR}^{in}+c_{kR}^{out}=\sqrt{u_{kR}}a_k,
\end{equation}
where $u_{kL}$ ($u_{kR}$) is the effective coupling parameter between the internal and
external fields of wavenumber $k$
at mirror $L$ ($R$). In the following we assume identical mirrors and the
coupling is independent of the wavenumber where
$u_{kL}=u_{kR}=u$. For example, in tapered nanofibers made of silica the tapered zones can be considered as effective mirrors \cite{Vetsch2010,Goban2012}. For nanophotonic waveguides, e.g. made of silicon, grating couplers are used to couple the internal-external fields that considered as effective mirrors \cite{Taillaert2006}.

Next we treat a ring waveguide. We assume an external nanofiber to be very close to a waveguide at a fixed point, as seen in figure (\ref{Ring}). Photons can tunnel between the waveguide and the nanofiber at this point. The input and output external fields
of wavenumber $k$ are given by $c_{k}^{in}$ and
$c_{k}^{out}$, which are related to waveguide mode $k$ by the boundary
condition \cite{Walls2008,Gardiner2010}
\begin{equation}
c_{k}^{in}+c_{k}^{out}=\sqrt{u_{k}}a_k,
\end{equation}
where $u_{k}$ is the coupling parameter between the internal and
external fields of wavenumber $k$
at the tunneling point that is proportional to the overlap between the waveguide and the nanofiber fields. In the following we assume the
coupling to be independent of the wavenumber where
$u_{k}=u$.

We introduce BIT into a system of a ring nanoscale waveguide, and BIO into a system of a linear nanoscale
waveguide, where BIT and BIO require three fields inside the waveguide, a signal photon, $(k,\omega_k)$, a pump photon, $(k-q,\omega_{k-q})$, and an acoustic phonon, $(q,\Omega_q)$. Therefore, we limit our discussion to the Hamiltonian
\begin{eqnarray} \label{QuadHam}
H&=&\hbar\omega_{k}\ a_k^{\dagger}a_k+\hbar\omega_{k-q}\ a_{k-q}^{\dagger}a_{k-q}+\hbar\Omega_q\ b^{\dagger}_qb_q \nonumber \\
&+&\hbar g^{\ast}\ b^{\dagger}_qa_{k-q}^{\dagger}a_k+\hbar g\ b_qa_k^{\dagger}a_{k-q}.
\end{eqnarray}

The first step towards treating BIT and BIO, in adapting the concept of
phonon-polaritons, is to bring the Hamiltonian into a quadratic form that includes
effective coupling between signal
photons and phonons that induced by SBS in the presence of the pump field. This procedure can be achieved in assuming the strong pump
field to be a classical one, where the obtained effective photon-phonon coupling is found to be related on the
pump field intensity. 

We apply two external
fields, a strong control field, $(\omega_p,k-q)$, that is described by $c_{k-q}^{in}$, in order to excite the waveguide pump field, and a weak probe field, $(\omega_s,k)$, that is described by $c_k^{in}$, in order to excited the
waveguide signal field. When the external probe field is tuned to resonance with the signal
waveguide field, we expect a strong penetration of the probe field into
the waveguide. Here SBS can take place within the waveguide due to the presence of the strong
pump field. Induced Brillouin scattering from the
signal field to the pump field assisted by an acoustic phonon will appear at the
required phase-matching. But a coherent superposition between the system eigenstates
inside the waveguide lead to a gap that open in the transmission spectrum and cause
a complete reflection of the probe field from the linear waveguide mirror, and a complete transmission for the case of a ring waveguide. In the
following we study the process in much details for the two cases of linear and ring waveguides. 

Let us concentrate now in the pump field at wavenumber $k-q$. The input pump field is represented by $c_{k-q}^{in}$. The pump field is enough strong in order to neglect any change in it due to SBS. Hence, using the input-output formalism \cite{Gardiner2010}, we get the equation of motion for the pump operator
\begin{equation}
\frac{d}{dt}\tilde{a}_{k-q}\approx-i\Delta_{k-q}\ \tilde{a}_{k-q}+\sqrt{u}\ \tilde{c}_{k-q}^{in},
\end{equation}
where $\Delta_{k-q}=\omega_{k-q}-\omega_p-i\left(u+\frac{\gamma}{2}\right)$, and we defined $a_{k-q}=\tilde{a}_{k-q}e^{-i\omega_pt}$, with $c_{k-q}^{in}=\tilde{c}_{k-q}^{in}e^{-i\omega_pt}$. The photon direct damping into free space is included phenomenologically through the damping rate $\gamma$. 

At steady state we have $\frac{d}{dt}\tilde{a}_{k-q}=0$. Then, for the pump field we get
\begin{equation} \label{pump}
a_{k-q}=\frac{\sqrt{u}}{i\Delta_{k-q}}c_{k-q}^{in}.
\end{equation}
We define $\hat{n}_{k-q}=a_{k-q}^{\dagger}a_{k-q}$, where $\hat{n}_{k-q}=\frac{u}{|\Delta_{k-q}|^2}\hat{n}_{k-q}^{in}$, with $\hat{n}_{k-q}^{in}=c_{k-q}^{in\dagger}c_{k-q}^{in}$. The average
value is $n_{k-q}=\langle a_{k-q}^{\dagger}a_{k-q}\rangle$, then $n_{k-q}=\frac{u}{|\Delta_{k-q}|^2}n_{k-q}^{in}$, with $n_{k-q}^{in}=\langle c_{k-q}^{in\dagger}c_{k-q}^{in}\rangle$. Note that $n_{k-q}^{in}$ has a unit of angular frequency.

\section{Phonon-Polaritons Point of View}

We start from the above Hamiltonian (\ref{QuadHam}) that includes only the three appropriate waveguide fields. The system can be linearized by assuming a strong classical pump
field, where the Hamiltonian casts into a quadratic one. We use the
previous results, where the pump field operator, at steady state in using
the input-output formalism, is given by equation (\ref{pump}). In the
classical limit we get
\begin{equation}
\beta_{k-q}=\frac{\sqrt{u}}{i\Delta_{k-q}}\alpha_{k-q},
\end{equation}
where $\beta_{k-q}=\langle a_{k-q}\rangle$, and
$\alpha_{k-q}=\langle c_{k-q}^{in}\rangle$. We achieve the quadratic Hamiltonian
\begin{equation}
H=\hbar\omega_{k}\ a_k^{\dagger}a_k+\hbar\Omega_q\ b^{\dagger}_qb_q+\hbar g^{\ast}\beta_{k-q}^{\ast}\ b^{\dagger}_qa_k+\hbar g\beta_{k-q}\ b_qa_k^{\dagger}.
\end{equation}
We move to a rotating frame that oscillates with the pump waveguide frequency $\omega_{k-q}$, in
applying the transformation $H\rightarrow U^{\dagger}HU-A$, where
$U=e^{-i\omega_{k-q}t\ a_k^{\dagger}a_k}$, and $A=\hbar\omega_{k-q}\ a_k^{\dagger}a_k$. We
get the Hamiltonian
\begin{equation}
H=\hbar\bar{\omega}_k\ a^{\dagger}_ka_k+\hbar\Omega_q\ b^{\dagger}_qb_q+\hbar f^{\ast}_{k-q}\ b^{\dagger}_qa_k+\hbar f_{k-q}\ b_qa^{\dagger}_k,
\end{equation}
where $\bar{\omega}_k=\omega_k-\omega_{k-q}$, with the effective coupling
\begin{equation}
f_{k-q}=\frac{g\sqrt{u}}{i\Delta_{k-q}}\alpha_{k-q}.
\end{equation}

Here we consider the strong coupling regime in which the effective coupling
$|f_{k-q}|$ is larger than the photon and phonon damping rates. The Hamiltonian can be diagonalized by introducing the polariton operators
\begin{equation}\label{PolTr}
A_{kq}^{\pm}=X_{kq}^{\pm}\ b_q+Y_{kq}^{\pm}\ a_k,
\end{equation}
which are coherent superposition of photons and phonons. The coherent
mixing amplitudes are
\begin{equation}
X_{kq}^{\pm}=\pm\sqrt{\frac{D_{kq}\mp\delta_{kq}}{2D_{kq}}},\ \ \ Y_{kq}^{\pm}=\frac{f_{k-q}^{\ast}}{\sqrt{2D_{kq}(D_{kq}\mp\delta_{kq})}},
\end{equation}
where
\begin{equation}
D_{kq}=\sqrt{\delta_{kq}^2+|f_{k-q}|^2},\ \ \ \delta_{kq}=\frac{\bar{\omega}_k-\Omega_q}{2}.
\end{equation}
We have the normalization condition $|X_{kq}^{\pm}|^2+|Y_{kq}^{\pm}|^2=1$. The polariton Hamiltonian reads
\begin{equation}
H=\sum_{\mu}\hbar\Omega_{kq}^{\mu}\ A_{kq}^{\mu\dagger}A_{kq}^{\mu},
\end{equation}
with the polariton dispersions
\begin{equation}
\Omega_{kq}^{\pm}=\frac{\bar{\omega}_k+\Omega_q}{2}\pm D_{kq}.
\end{equation}

\begin{figure}
\includegraphics[width=0.8\linewidth]{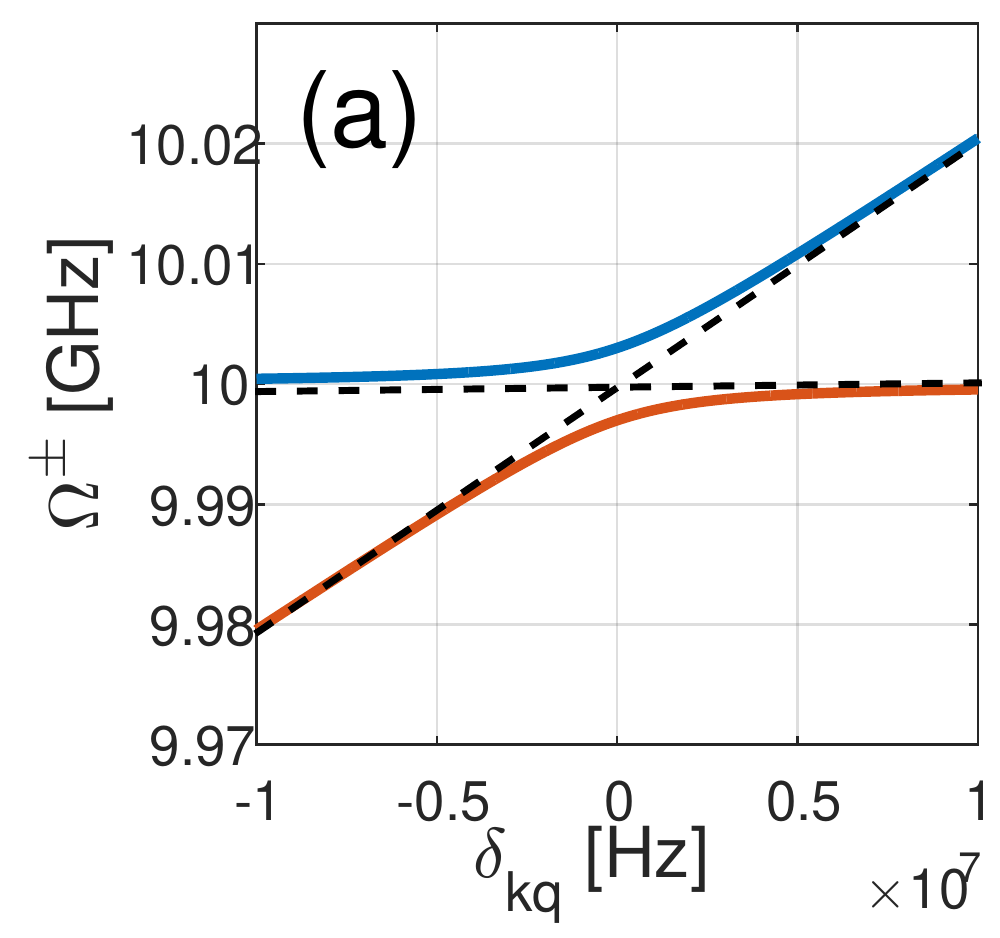}

\includegraphics[width=0.8\linewidth]{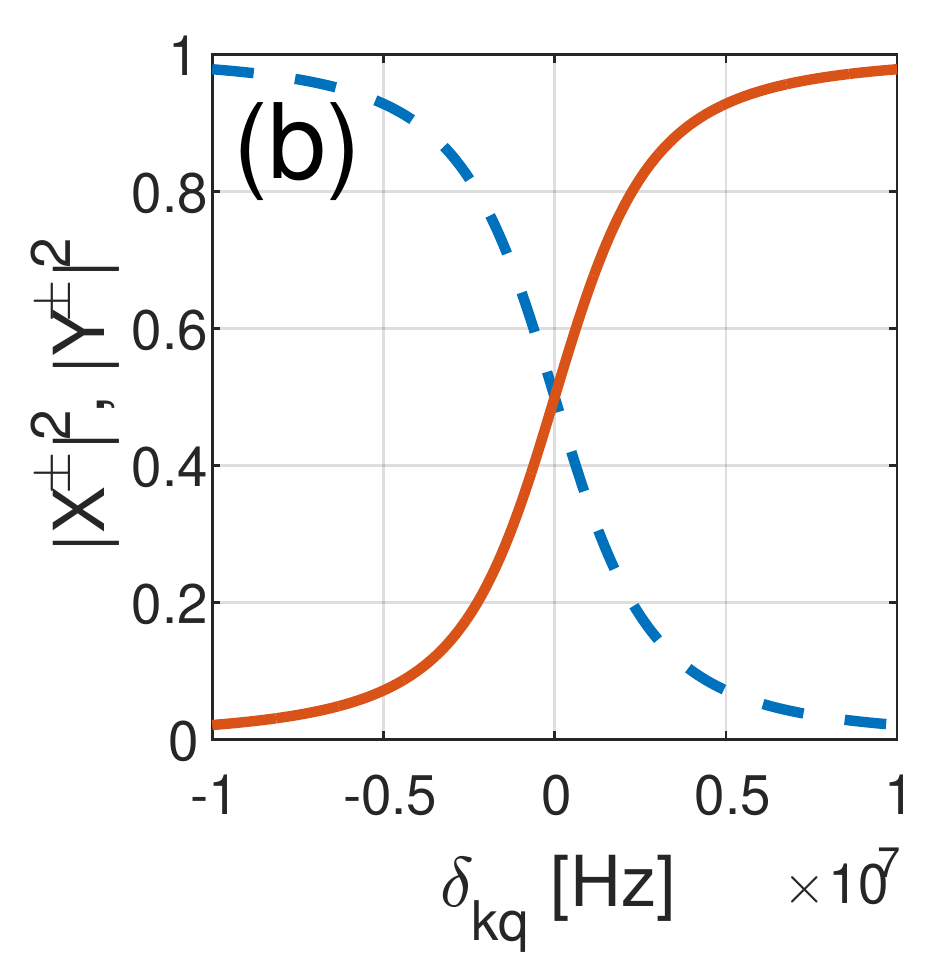}
\caption{(a) The upper and lower polariton frequencies, $\Omega^{\pm}_{kq}$, as a function of the detuning $\delta_{kq}$. The horizontal dashed line is for the fixed acoustic phonon, and the linear dashed line is of the photon mode. (b) The photon and phonon fractions $|X_{kq}^{\pm}|^2$ and $|Y_{kq}^{\pm}|^2$ as a function of $\delta_{kq}$. For the lower polariton branch, the dashed line is the photonic fraction and the solid line for the phononic fraction, and the opposite for the upper polariton branch.}
\label{PolDis}
\end{figure}

We illustrate the polariton modes using typical nanoscale waveguide numbers. We consider a fixed phonon mode of frequency $\Omega_q=10$~GHz, and change the scaled signal frequency $\bar{\omega}_k=\omega_k-\omega_p$, where our control frequency is $\delta_{kq}=(\bar{\omega}_k-\Omega_q)/2$ for a fixed pump frequency (in using external-internal resonance pump frequency of $\omega_p=\omega_{k-q}$). For the photon-phonon coupling we use $f=3$~MHz (to neglect the dependent on wavenumbers in this region). In figure (\ref{PolDis}.a) we plot the upper and lower polariton dispersion. The plot represents the polariton frequencies, $\Omega^{\pm}$ (relative to the fixed pump frequency $\omega_p$), as a function of the detuning, $\delta_{kq}$, where we change $\bar{\omega}_k$ relative to a fixed $\Omega_q$. Around the intersection point, at $\bar{\omega}_k=\Omega_q$, the photon and phonon modes are mixed and split to give two polariton modes, with rabi splitting frequency of $2|f|$. In figure (\ref{PolDis}.b) we plot the photon and phonon factors in the formation of polaritons. Namely, we plot $|X_{kq}^{\pm}|^2$ and $|Y_{kq}^{\pm}|^2$ as a function of $\delta_{kq}$. At large negative detuning the lower polariton branch becomes photonic with $|Y^-|^2\approx 1$ and $|X^-|^2\approx 0$, and the upper polariton branch becomes phononic with $|Y^+|^2\approx 0$ and $|X^+|^2\approx 1$. The opposite for large positive detuning, the lower polariton branch becomes phononic with $|Y^-|^2\approx 0$ and $|X^-|^2\approx 1$, and the upper polariton branch becomes photonic with $|Y^+|^2\approx 1$ and $|X^+|^2\approx 0$. Around resonance the upper and lower polariton branches are significant mix of photons and phonons. At $\delta_{kq}\approx 0$ the polaritons are half photonic and half phononic, that is $|X^{\pm}|^2=|Y^{\pm}|^2\approx 1/2$.

The photon and phonon damping rates can be included
phenomenologically by the replacement $\omega_k\rightarrow\omega_k-i\gamma/2$
and $\Omega_q\rightarrow\Omega_1-i\Gamma/2$, where $\gamma$ is the photon damping rate, and $\Gamma$ is the phonon damping rate. This replacement leads to the
complex polariton frequencies $\bar{\Omega}_{kq}^{\pm}$, which can be
approximated in the strong coupling regime by
$\bar{\Omega}_{kq}^{\pm}=\Omega_{kq}^{\pm}-i\Gamma_{kq}^{\pm}$, with the
effective polariton damping rate
\begin{equation}
\Gamma_{kq}^{\pm}=|X_{kq}^{\pm}|^2\frac{\Gamma}{2}+|Y_{kq}^{\pm}|^2\frac{\gamma}{2}.
\end{equation}

\subsection{BIO in Linear Nanoscale Waveguides}

We present here the input-output formalism in the frame of the
polariton picture. For the linear waveguide, that is represented in figure (\ref{Linear}), the electromagnetic fields at the left and right sides of
the waveguide are given by
\begin{eqnarray}
H_R&=&\sum_k\int
d\omega_k^{\prime}\ \hbar\omega_k^{\prime}\ c_R^{\dagger}(\omega_k^{\prime})c_R(\omega_k^{\prime}), \nonumber \\
H_L&=&\sum_k\int
d\omega_k^{\prime}\ \hbar\omega_k^{\prime}\ c_L^{\dagger}(\omega_k^{\prime})c_L(\omega_k^{\prime}),
\end{eqnarray}
with the interaction Hamiltonians
\begin{eqnarray}
V_R&=&\sum_k\int
d\omega_k^{\prime}\ i\hbar
\ u_R(\omega_k^{\prime})\left(c_R^{\dagger}(\omega_k^{\prime})a_k-a_k^{\dagger}c_R(\omega_k^{\prime})\right), \nonumber \\
V_L&=&\sum_k\int
d\omega_k^{\prime}\ i\hbar
\ u_L(\omega_k^{\prime})\left(c_L^{\dagger}(\omega_k^{\prime})a_k-a_k^{\dagger}c_L(\omega_k^{\prime})\right), \nonumber \\
\end{eqnarray}
where the coupling parameters are $u_R(\omega_k^{\prime})$ and $u_L(\omega_k^{\prime})$. Using the inverse transformation of (\ref{PolTr}), the photon operator can be
written in term of the polariton operators by
\begin{equation} \label{Inverse}
a_k=\sum_{\mu}Y_{kq}^{\mu\ast}\ A_{kq}^{\mu}.
\end{equation}
The equations of motion for the polariton operators, considering coupling at the left and
right mirrors, are
\begin{eqnarray}
\frac{d}{dt}A_{kq}^{\mu}&=&-i\bar{\Omega}_{kq}^{\mu}\ A_{kq}^{\mu}+\sqrt{u}Y_{kq}^{\mu}\ c^{in}_k-uY_{kq}^{\mu}\ a_{k}, \nonumber \\
\frac{d}{dt}A_{kq}^{\mu}&=&-i\bar{\Omega}_{kq}^{\mu}\ A_{kq}^{\mu}-\sqrt{u}Y_{kq}^{\mu}\ c^{out}_k+uY_{kq}^{\mu}\ a_{k},
\end{eqnarray}
where $c^{in}_k=c^{in}_{Lk}+c^{in}_{Rk}$, and
$c^{out}_k=c^{out}_{Lk}+c^{out}_{Rk}$. We assumed two identical mirrors with
coupling parameter $u$. We have the boundary conditions
\begin{equation}
\sqrt{u}\ a_k=c^{in}_{Lk}+c^{out}_{Lk},\ \ \ \sqrt{u}\ a_k=c^{in}_{Rk}+c^{out}_{Rk}.
\end{equation}
The incident probe field is taken only from the left side, then $c^{in}_k=c^{in}_{Lk}$,
and $c^{in}_{Rk}=0$.

Taking Fourier transform of the equations into $\omega$-space, we
get the equations
\begin{eqnarray} \label{PolEqs}
i(\bar{\Omega}_{kq}^{\mu}-\omega)\tilde{A}_{kq}^{\mu}=\sqrt{u}Y_{kq}^{\mu}\ \tilde{c}^{in}_{Lk}-uY_{kq}^{\mu}\ \tilde{a}_{k}, \nonumber \\
i(\bar{\Omega}_{kq}^{\mu}-\omega)\tilde{A}_{kq}^{\mu}=-\sqrt{u}Y_{kq}^{\mu}\ (\tilde{c}^{out}_{Lk}+\tilde{c}^{out}_{Rk})+uY_{kq}^{\mu}\ \tilde{a}_{k}. \nonumber \\
\end{eqnarray}
The first equation, using (\ref{Inverse}), yields
\begin{equation}
\tilde{a}_{k}=-i\frac{\sqrt{u}\Lambda_{kq}}{1-iu\Lambda_{kq}}\tilde{c}^{in}_{Lk},
\end{equation}
where
\begin{equation}
\Lambda_{kq}=\sum_{\mu}\frac{|Y_{kq}^{\mu}|^2}{\bar{\Omega}_{kq}^{\mu}-\omega}.
\end{equation}
Using the boundary conditions
\begin{equation}
\sqrt{u}\ \tilde{a}_k=\tilde{c}^{in}_{Lk}+\tilde{c}^{out}_{Lk},\ \ \ \sqrt{u}\ \tilde{a}_k=\tilde{c}^{out}_{Rk},
\end{equation}
we get the reflection complex amplitude
\begin{equation}
r=|r|e^{i\phi_r}=\frac{\tilde{c}^{out}_{Lk}}{\tilde{c}^{in}_{Lk}}=\frac{-1}{1-iu\Lambda_{kq}},
\end{equation}
and the transmission complex amplitude
\begin{equation}
t=|t|e^{i\phi_t}=\frac{\tilde{c}^{out}_{Rk}}{\tilde{c}^{in}_{Lk}}=-i\frac{u\Lambda_{kq}}{1-iu\Lambda_{kq}}.
\end{equation}
The reflection parameter is
\begin{equation}
R=\frac{\langle\tilde{c}^{out\dagger}_{Lk}\tilde{c}^{out}_{Lk}\rangle}{\langle\tilde{c}^{in\dagger}_{Lk}\tilde{c}^{in}_{Lk}\rangle}=\frac{1}{|1-iu\Lambda_{kq}|^2},
\end{equation}
and the transmission parameter is
\begin{equation}
T=\frac{\langle\tilde{c}^{out\dagger}_{Rk}\tilde{c}^{out}_{Rk}\rangle}{\langle\tilde{c}^{in\dagger}_{Lk}\tilde{c}^{in}_{Lk}\rangle}=\frac{u^2|\Lambda_{kq}|^2}{|1-iu\Lambda_{kq}|^2},
\end{equation}
using $T+R+A=1$ gives the absorption parameter
\begin{equation}
A=\frac{iu(\Lambda_{kq}^{\ast}-\Lambda_{kq})}{|1-iu\Lambda_{kq}|^2}.
\end{equation}

We illustrate the results using typical numbers for nanoscale
waveguides. The photon damping rate is $\gamma=10^5$~Hz, the phonon damping
rate is $\Gamma=10^6$~Hz, the mirror coupling is $u=10^{6}$~Hz, and the photon-phonon
coupling is $g=10^4$~Hz. The signal photon mode at wavenumber $k$ is taken to be
$\omega_k=10^{14}$~Hz, the phonon mode at $q$ is $\Omega_q=10^{10}$~Hz, and the pump photon mode at wavenumber $k-q$ is taken to be
$\omega_{k-q}=\omega_k-\Omega_q$.

\begin{figure}
\includegraphics[width=0.8\linewidth]{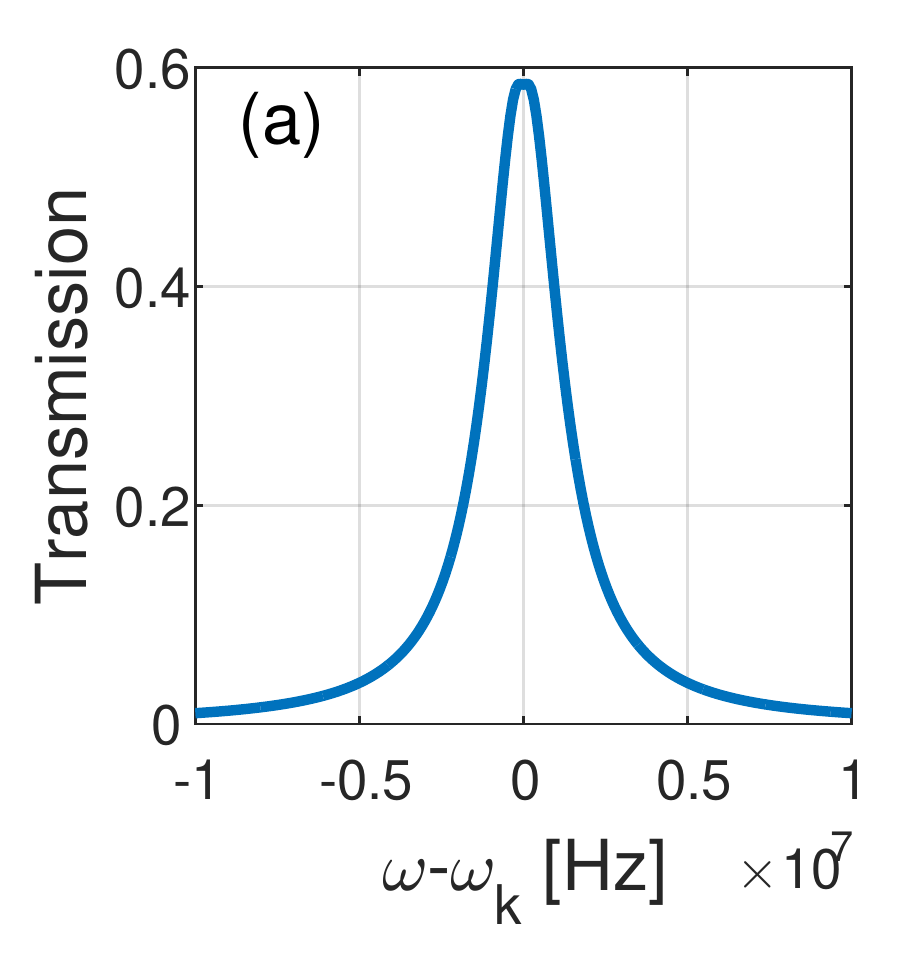}

\includegraphics[width=0.8\linewidth]{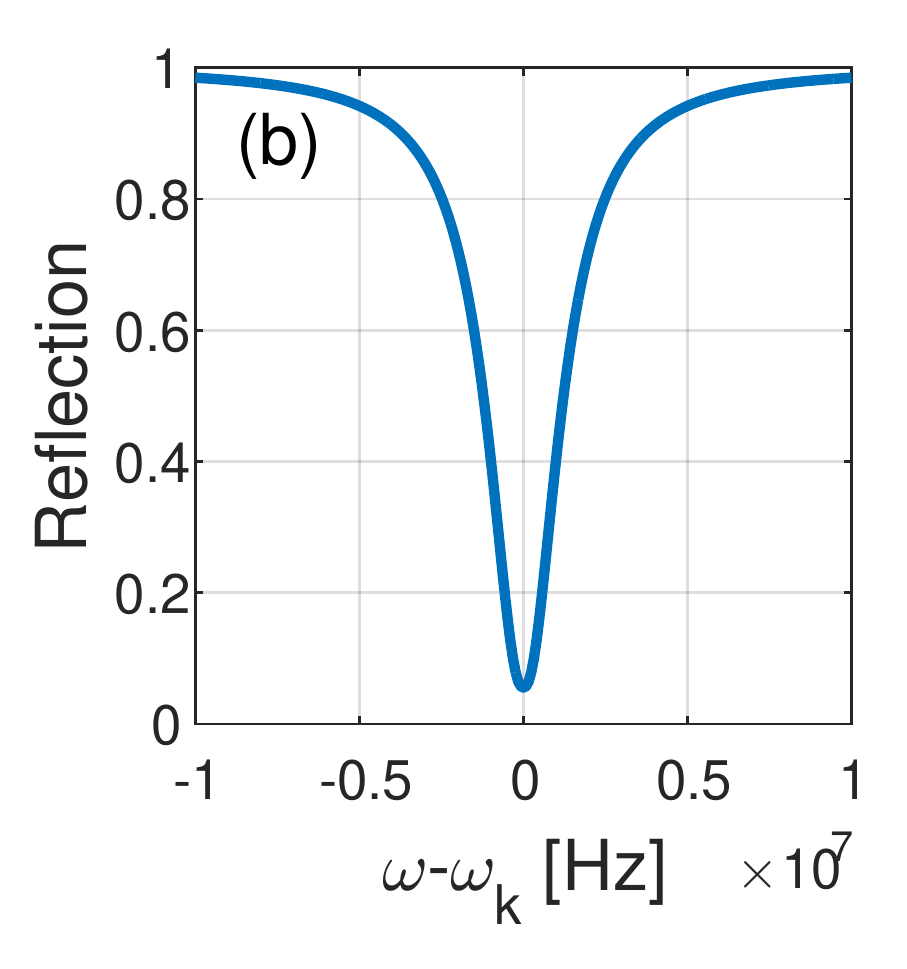}
\caption{Linear waveguide. (a) The signal transmission $T$ vs. frequency $\omega-\omega_k$. (b) The signal reflection $R$ vs. frequency $\omega-\omega_k$. The pump is at resonance, $\omega_p=\omega_{k-q}$, at low intensity of $n_{k-q}^{in}=10^{8}$~Hz.}
\label{TR2}
\end{figure}

We present first the case of small pump intensity of $n_{k-q}^{in}=10^{8}$~Hz, and for resonance external-internal pump, that is $\omega_p=\omega_{k-q}$. We present the signal transmission spectrum in figure (\ref{TR2}.a), and the signal reflection spectrum in figure (\ref{TR2}.b), as a function of the detuning frequency $(\omega-\omega_k)$. We are in the weak coupling regime, hence one peak appears in the transmission, and one dip in the reflection, at the signal field frequency, that is at $\omega=\omega_k$. The concept of polaritons is irrelevant here, where full transmission and small reflection appear around resonance.

\begin{figure}
\includegraphics[width=0.8\linewidth]{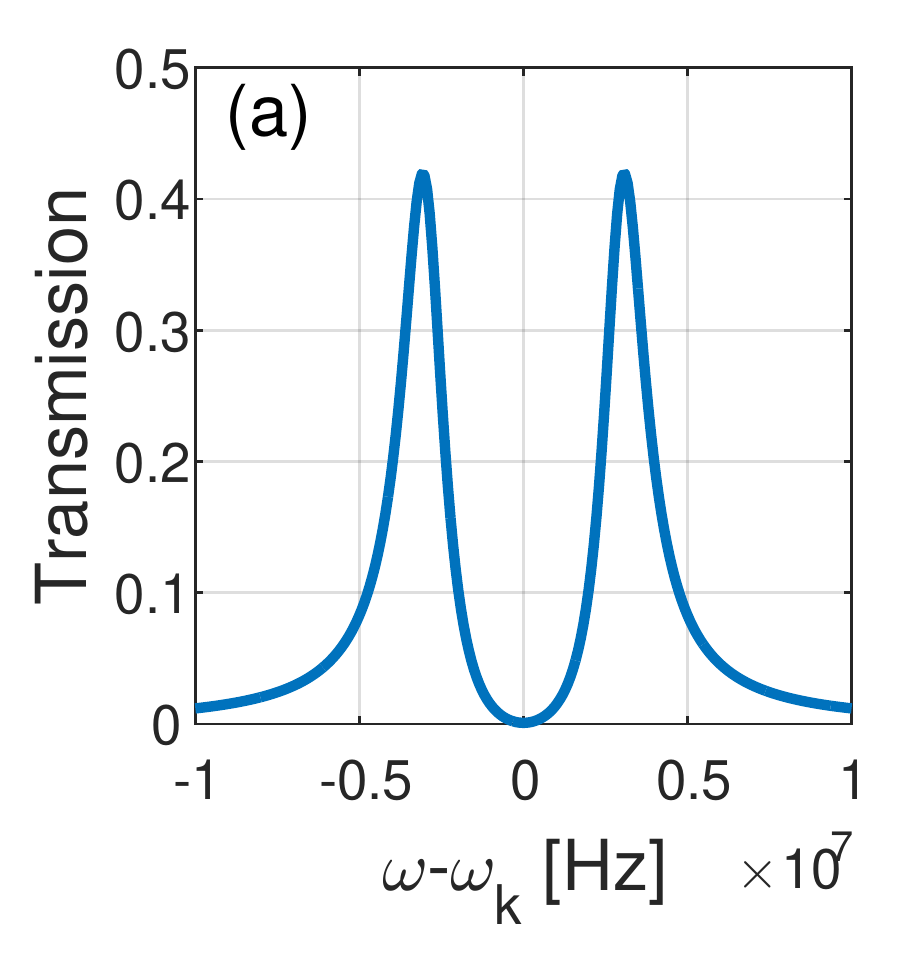}

\includegraphics[width=0.8\linewidth]{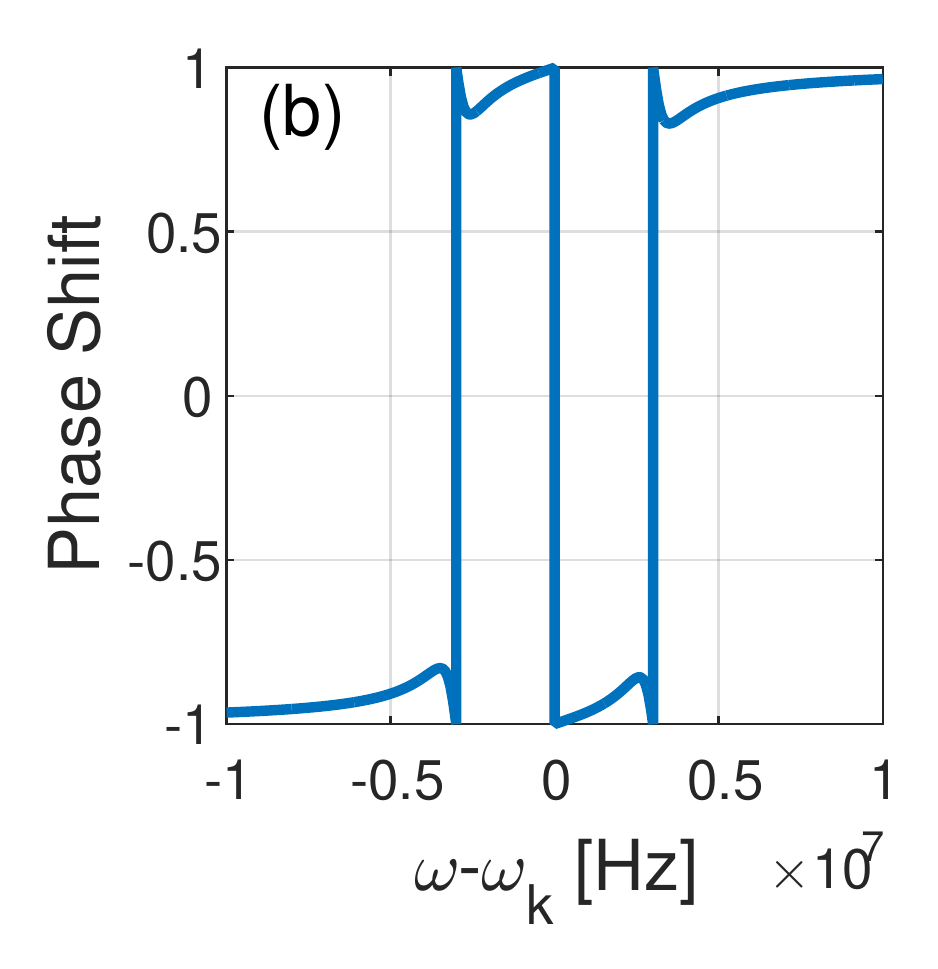}
\caption{Linear waveguide. (a) The signal transmission $T$ vs. frequency $\omega-\omega_k$. (b) The signal transmission phase
  shift $\phi_t$ vs. frequency $\omega-\omega_k$. The pump is at resonance, $\omega_p=\omega_{k-q}$, of large intensity $n_{k-q}^{in}=10^{11}$~Hz.}
\label{TS1}
\end{figure}

\begin{figure}
\includegraphics[width=0.8\linewidth]{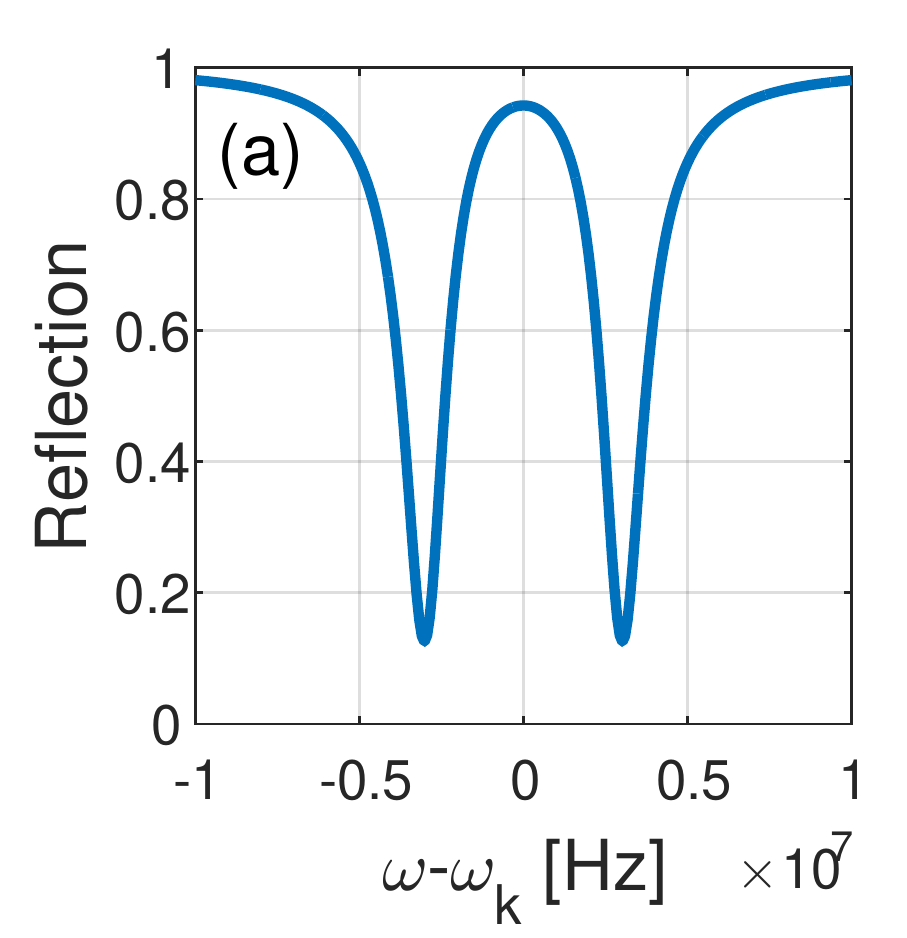}

\includegraphics[width=0.8\linewidth]{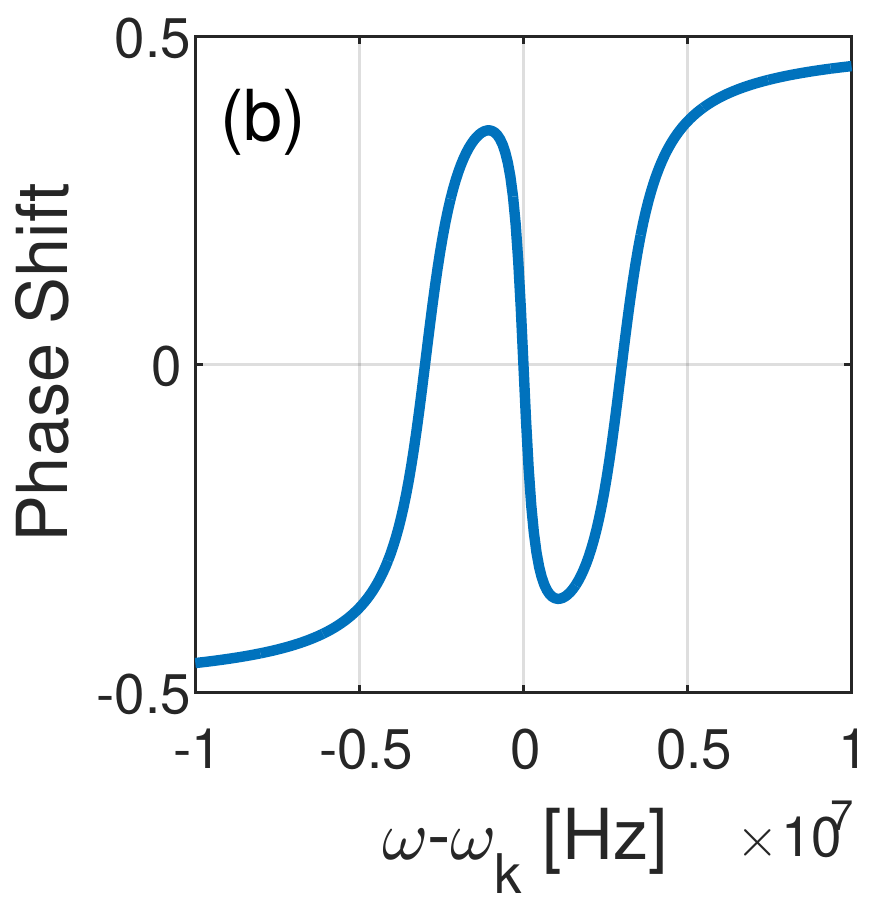}
\caption{Linear waveguide. (a) The signal reflection $R$ vs. frequency $\omega-\omega_k$. (b) The signal reflection phase
  shift $\phi_t$ vs. frequency $\omega-\omega_k$. The pump is at resonance, $\omega_p=\omega_{k-q}$, of large intensity $n_{k-q}^{in}=10^{11}$~Hz.  For a ring waveguide the plots represent the transmission and its phase shift of the signal field through the external nanofiber.}
\label{RS1}
\end{figure}

Next we increase the pump intensity in order to achieve the strong coupling regime keeping the case of external-internal resonance at $\omega_p=\omega_{k-q}$. We use the pump intensity around $n_{k-q}^{in}=10^{11}$~Hz. We plot the signal
transmission $T$ in figure (\ref{TS1}.a), and its phase shift in figure (\ref{TS1}.b), as a
function of the frequency $\omega-\omega_q$, for the case of resonance pump, that is $\omega_p=\omega_{k-q}$. No transmission appears now around resonance. We plot the signal
reflection $R$ in figure (\ref{RS1}.a), and its phase shift in figure (\ref{RS1}.b), as a
function of the frequency $\omega-\omega_q$. Here, complete reflection appears around resonance and we get BIO.

The opacity
appears as the photons and phonons coherently are mixed and split to give two
polariton modes that are separated at the intersection point by a frequency splitting of twice the
coupling parameter (Rabi frequency). Without the pump field, the external probe field of
frequency $\omega_s$ has
a resonance with a waveguide mode at $\omega_k$. The pump field removes this
resonance through SBS, and new two resonances appear at the polariton frequencies. We conclude that the real excitations in the strong coupling regime are polaritons and not independent photons and phonons.

\subsection{BIT in Ring Nanoscale Waveguides}

We consider the case of a one dimensional nanoscale waveguide that forms
a ring, which is represented in figure (\ref{Ring}). The treatment of the previous section holds here, as we used periodic boundary conditions for the propagating light inside the linear waveguide. Now
we remove the edge mirrors and connect the two sides to form a ring. The
input-output coupling is achieved by an external optical nanofiber that coupled
to the waveguide ring at a fixed point.

In terms of polaritons we get the same system of equations as in eqs.(\ref{PolEqs}), where the input and output fields obey now the boundary condition
\begin{equation}
\sqrt{u}\ a_k=c^{in}_{k}+c^{out}_{k}.
\end{equation}
Repeating the calculations for a ring waveguide, we get the complex amplitude
\begin{equation}
t=|t|e^{i\phi_r}=\frac{\tilde{c}^{out}_{k}}{\tilde{c}^{in}_{k}}=\frac{-1}{1-iu\Lambda_{kq}},
\end{equation}
with the transmission parameter
\begin{equation}
T=\frac{\langle\tilde{c}^{out\dagger}_{k}\tilde{c}^{out}_{k}\rangle}{\langle\tilde{c}^{in\dagger}_{k}\tilde{c}^{in}_{k}\rangle}=\frac{1}{|1-iu\Lambda_{kq}|^2},
\end{equation}
and using $T+A=1$ in order to get the absorption parameter. The present $t$ is identical to $r$ in the previous case of a linear
waveguide, where all the above plots of $r$, in figure (\ref{RS1}), are the same for the present $t$. Hence for a ring waveguide we get BIT for the light propagating in the external fiber, rather than BIO in the above case of linear waveguides. Namely, in the absence of the pump field the probe field is strongly coupled to the waveguide signal field and only small part transmitted along the fiber. But once the pump is switched on the fiber-waveguide resonance at the probe frequency blocked and the probe light continue along the fiber without entering the waveguide. Namely the system becomes transparent for the probe field.

\section{Conclusions}

We adopt the concept of phonon-polariton in order to explain the phenomena of BIO in linear waveguides and BIT in ring waveguides. We show how the strong coupling regime among a phonon and a photon can be achieved through SBS in the presence of a strong pump field. Coherent oscillations between the phonon and photon fields inside a waveguide is possible once the effective phonon-photon coupling becomes larger than their damping rates. This regime is reachable for nanoscale waveguides, in which radiation pressure dominates over conventional electrostriction, and for long lived phonons within the waveguide.

The input-output method provides us with a tool for observing the system collective eigenmodes, by coupling the internal waveguide states to the external radiation field through the effective mirrors in linear waveguides and through a nearby nanofiber in ring waveguides. The method yields linear optical spectra and phase shifts for an external probe field, where the intensity of the pump field serves us as a control parameter. In a linear waveguide at low pump intensity a single peak appears in the transmission spectrum and a single dip in the reflection spectrum. But for high pump intensity, once the strong coupling regime is achieved, two peaks appear in the transmission spectrum and two dips in the reflection spectrum, which are a signature for the formation of phonon-polaritons. Namely, at a frequency where a transparency is expected in a linear waveguide an opacity obtained due to SBS. Similar consideration hold for a ring waveguide, but in the strong coupling where an opacity is expected a transparency is obtained.

The advantage of nanoscale waveguides is in their easy integration into all-optical on-chip platform. BIT and BIO introduced in the present paper can be of importance for quantum information processing and communication involving photons and phonons. BIT and BIO can be tailored for realizing phonon and photon Brillouin laser. Phonon-polaritons, as exotic states made of a coherent mix of light and mechanical excitations, are a promising candidate for the implementation of non-classical and entangled states.

\section*{Acknowledgment}

The author thanks Klemens Hammerer for very fruitful discussions.

\end{document}